\documentstyle[12pt, epsf]{article}
\begin{document}
\newcommand{\beq}{\begin{equation}}
\newcommand{\eeq}{\end{equation}}
\newcommand{\be}{\begin{equation}}
\newcommand{\ee}{\end{equation}}
\newcommand{\lll}{\lambda}
\def\Journal#1#2#3#4{{#1} {\bf #2}, #3 (#4)}


\def\st{\scriptstyle}
\def\sst{\scriptscriptstyle}
\def\mco{\multicolumn}
\def\epp{\epsilon^{\prime}}
\def\vep{\varepsilon}
\def\ra{\rightarrow}
\def\al{\alpha}
\def\ab{\bar{\alpha}}
\def\be{\begin{equation}}
\def\ee{\end{equation}}
\def\bea{\begin{eqnarray}}
\def\eea{\end{eqnarray}}
\def\CPbar{\hbox{{\rm CP}\hskip-1.80em{/}}}

\begin{flushright}
\begin{tabular}{l}
ITEP-TH-22/06
\end{tabular}
\end{flushright}

\vskip1cm

\centerline{\large
\bf Spontaneous Creation of the  Brane World and }
\bigskip
\centerline{\large
\bf  Direction of the Time Arrow
}

\vspace{0.5cm}

\centerline{\sc  A.S. Gorsky}

\vspace{5mm}

\centerline{\it Institute of Theoretical and Experimental Physics }
\centerline{\it  B. Cheremushkinskaya ul. 25, 117259 Moscow, Russia}

\def\thefootnote{\fnsymbol{footnote}}%
\vspace{1.5cm}

\centerline{\bf Abstract}
In this note we   consider the spontaneous creation of the
brane world in   five-dimensional space with
nondynamical external four-form field via  spherically asymmetric
bounce solution.
We argue that spherically asymmetric bounce suggests several inequivalent
directions of the time arrow upon the analytic continuation
to the space-time with Lorentzian signature. It it shown that S-branes
in the imaginary time emerge naturally upon the particular continuation.
\vspace{1.0cm}


${\bf {1}}$. \quad The notion of  wave function of the universe \cite{hh} is
important
in the quantum cosmology. It depends on the metric and possible
matter fields at
3d space slice and can be obtained via the standard quantum mechanical
rules via summation over all geometries with  particular
3d boundary.
There are several proposals and prescriptions concerning its
actual calculation \cite{hh,vilenkin,linde} however this issue
is still under debates since "tunneling from nothing" scenario
of Vilenkin and "no-boundary" prescription of Hartle-Hawking
yield different exponential factors at the quasiclassical regime.
This wave function should obey Wheeler-De Witt equation where
the scale factor plays the role of canonical coordinate
in the minisuperspace approximation.

This issue can be tested in the brane world scenario \cite{rubakov1}
implying the localization of the gravity on the brane \cite{ran}
(see \cite{rew} for a review). Similar to
the "tunneling from nothing" scenario it is natural to consider
the spontaneous creation of the brane world. Being the solitonic object from five-dimensional point of view
four-dimensional brane world can be produced nonperturbatively in a
higher dimensional generalization of the Schwinger process.
The creation of the extended objects in the external flux
has been previously  used to reduce cosmological constant
\cite{bt} and was generalized for the case of the multiple fluxes
in \cite{bp}.

The explicit  examples
of the nonperturbative production of the
brane world deal with the  Schwinger type
process in the external nondynamical four-form field in
five dimensions \cite{gs1,gs2} or with the production of the
bubble of the true vacuum  in the false vacuum
in the theory with additional scalar field and peculiar
potential \cite{garriga}. The wave function of the universe
can be treated in this way more literally since it
corresponds to the wave function of the brane
degrees of freedom \cite{vilenkin2} and metric
on the brane is fixed classically by the geometry of embedding.

It shall be important in what follows that bounces found
in \cite{gs1,gs2} are spherically asymmetric and
have several turning points. Recently gravitational spherically asymmetric
bounces were considered
as well \cite{tye,brushtein,bk}.
The  asymmetry of gravitational bounces  emerges if the quantum
radiation during  tunneling or before tunneling
is taken into account. It turns out that in this case
the emerging wave function of the universe depending
on the scale factor predicts the finite nonvanishing value of the
cosmological constant.

To define the wave function of the universe
one has to select the direction of the time arrow. In the
spontaneous creation of universe via tunneling process
there is no initial prescription of the time arrow which is fixed by
the surface where continuation from the Euclidean to the Lorentzian
geometries takes place. The general conditions imposed on these
surfaces have been discussed in \cite{hartle}.

In this note we shall discuss the inequivalent continuations into the Lorentzian space-time
of the spherically asymmetric Euclidean bounce solution in the external
four-form  field in five dimensions yielding the spontaneous
creation of the brane world.
Our solution involves simultaneously charged and neutral
branes.
It turns out that spherically asymmetric bounces allow several inequivalent
continuations with different brane content upon the
materialization. That is spherically asymmetric Euclidean solution amounts to
several wave functions of the universe or directions of the time arrow. One
choice of the continuation surface amounts to the multiple neutral
branes and single charged brane upon the materialization and time arrow
turns out to be  parallel
to the worldvolume of the neutral branes  which are at rest in the
external field. This type of continuation resembles the process
of the particle creation during the tunneling \cite{rubakov}
however in our model example the  branes  created are involved into the asymmetric
bounce solution and strongly deform its shape.

Another continuation surface yields only  spherical charged brane in the
final state however the corresponding wave function of the brane universe
involves the summation over the array of the neutral branes. This
continuation surface provides the different direction of the time arrow and
neutral branes are identified with the  S-branes
introduced in \cite{gutperle}. Such array of S-branes in imaginary time has been related with
the stringy mode of the decaying brane in \cite{rastelli,msy,llm}
that is spontaneous creation of the brane world can be
identified with some decaying neutral brane. The brane world is
now defined on the charged branes with four-dimensional
De Sitter worldvolume geometry.

Note that the possibility of  two different time directions
induced by  continuation of the
two-dimensional Euclidean solution has
been mentioned in \cite{ovv} where the derivation of the
Hartle-Hawking wave function in the framework of  topological strings
has been discussed.
It was applied there to  explanation of   the relation between
the entropy of the black holes and partition functions
of the topological strings and was treated as some version
of open-closed duality. More recently similar issue
has been addressed in D=4 context in \cite{mac}.

This paper is organized as follows. First we shall describe
two examples of the spherically asymmetric  bounces responsible for the
Schwinger type processes. Then we discuss  continuation surfaces
corresponding to the different wave functions of the universe
and emphasize the essential role of S-branes. We conclude
with some speculations and open questions.

${\bf {2}}$. \quad To start with let us discuss  spherically asymmetric
Euclidean bounce
relevant for the Schwinger pair production in two dimensions. In
two dimensional electrodynamics the gauge field is nondynamical however
it provides  the nonperturbative pair production. The
leading contribution to the probability follows from the
Euclidean bounce solution which is just circle in  two
dimensional Euclidean space and represents the worldline of the
electron-positron pair. The probability of the process at
the exponential accuracy reads as
\beq
w_0\propto exp(-\frac{\pi m^2}{eE})
\eeq
where $m$ is the particle mass and $E$ is electric field.
However one can consider more complicated asymmetric bounce solutions
involving neutral bound states. The examples of such bounce are
shown at Fig.1. It is important to emphasize that in  nonperturbative pair production
process  the direction of the time arrow is fixed
from the very beginning and the continuation surface from the Euclidean to Lorentzian
space-time
should be transverse to this fixed direction. That is  orientation
of the spherically asymmetric bounce in the Euclidean space matters.
\begin{figure}
\epsfxsize=10cm
\centerline{\epsfbox{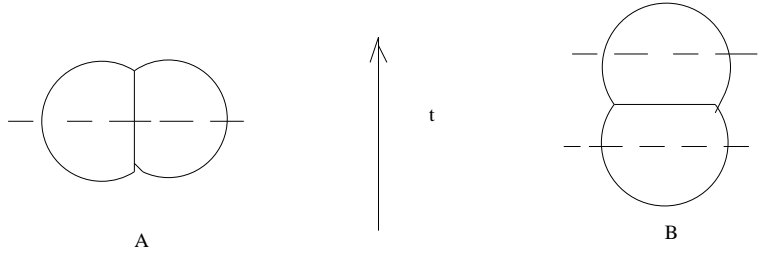}}
\caption{\small Two different orientations of
the spherically asymmetric Euclidean bounce responsible for the Schwinger
pair production in two dimensions.
}
\label{fig:122}
\end{figure}

As the basic example
we consider two different orientations of the same bounce at Fig.1
which result in the different final states upon materialization.
The action calculated on this bounce configuration reads as \cite{gs1}
\beq
S=\frac{2m^2}{eE}(\pi - arcsin(1-\frac{M^2}{4m^2})^{1/2})+ \frac{2Mm}{eE}(1-\frac{M^2}{4m^2})^{1/2}
\eeq
where $m(M)$ is the mass of the charged(neutral) particle.

With the  continuation surface  at Fig.1b we get  electron-positron pair
in the final state
that is  asymmetric bounce amounts to the nonperturbative correction to the
contribution from the spherically symmetric bounce. The neutral state
corresponds to the S-string localized in the
Euclidean time. One can easily consider more complicated bounce involving
many S-strings that is generic probability of the pair production at the exponential
accuracy reads as
\beq
W=w_0 +w_1+w_2 +\dots
\eeq
If the mass of the bound state is close to $2m$  the probability
involving $n$ S-strings can be approximated by $exp(-nS_0)$ and summation
over the arbitrary number of S-strings provides the result indicating
a kind of  thermal behavior
\beq
W_{\infty}\propto \frac{e^{-S_0}}{1-e^{-S_0}}
\eeq
The effective "temperature" of (0+1) dimensional  universe coincides with the Unruh-like temperature
for the particle moving with  constant acceleration $T=\frac{a}{2\pi}=\frac{eE}{2m\pi}$
in the external electric field.

With the different continuation surface at Fig.1a we get charged
pair and neutral bound state upon materialization that is different
final state which
does not exist  for the spherically symmetric bounce at all. The bubble
involving multiple neutral states
presented at Fig.2 yields with the  continuation cut A
the charged $e^{+}e^{-}$
pair and multiple neutral bound states in the final state.
Note that although the asymmetric bounce in the
Schwinger pair production   and the one responsible
for the creation of the brane world look similar there is
important difference. In the Schwinger production the time direction
is fixed while in the brane world creation time is defined
upon the continuation surface is chosen.

\begin{figure}
\epsfxsize=10cm
\centerline{\epsfbox{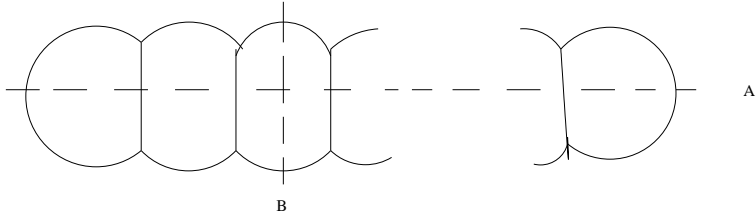}}
\caption{\small  Euclidean bounce involving the
large number of neutral branes.
}
\label{fig:125}
\end{figure}

More general situation concerns two simultaneous cuts at Fig.1b and
has the different interpretation. With two cuts we get  function
depending on two radii $\rho(R_1,R_2)$ which can be naturally
treated as the density matrix similar to case of asymmetric
gravitational bounce  in \cite{bk}. The example
represented at Fig.1b corresponds to the density matrix
$\rho(R,R)$ since both arguments correspond to the same  turning point.
In the limit  $M=2m$ the density matrix reduces to the pure
state case since the bounce degenerates to two copies of the spherically symmetric
bounce. One could also glue bounce along two cuts or two
bounces to get the bounce with toroidal geometry
corresponding to the thermal behaviour.

${\bf {3}}$. \quad Let us briefly review  spherically asymmetric
Euclidean bounce
solutions providing the spontaneous creation of the
brane world found in \cite{gs1,gs2}. The simplest model
\cite{gs1} involves  asymptotically flat five-dimensional
Euclidean space. It is assumed that negative cosmological constant
$\Lambda$ is exactly compensated by the positive energy
density from the field strength $H=dB$ of the four-form field B
which is nondynamical in five dimensions
\beq
\Lambda +\frac{kh_{+}^2}{2}=0
\eeq
where $k$ is five-dimensional gravitational constant and
$h_{+}$ is the asymptotic value of the Hodge dual of $H$. We
assume the existence of 3-branes charged with respect to
the external four-form field which can be glued together
along some junction manifolds.

The bounce configuration  is the generalization of
spherically symmetric bounce responsible for the Schwinger type
production of branes discussed in \cite{bt}. The asymmetric
bounce involves two spherical segments
of the charged branes
glued along  the circle junction with the neutral brane with
the disc topology which is higher dimensional
analogue of Fig.1. The external field jumps across
the charged branes and yields the effective cosmological
constant inside the bounce
\beq
\Lambda_{eff}= \frac{k(h_{-}^2 -h_{+}^2)}{2}
\eeq
with the corresponding radius of $AdS_5$ geometry
\beq
R^{2}_{AdS_5}= -\frac{24}{\Lambda_{eff}}
\eeq
The metric to the right of the neutral brane inside the bubble
reads as
\beq
ds^2=\frac{dz^2+d\rho^2 +\rho^2
d\Omega_3^2}{(1-\frac{(z-a)^2+\rho^2}{R_{ads}^2})^2}
\eeq
while the metric to the left of the neutral brane inside
the bubble is
\beq
ds^2=\frac{dz^2+d\rho^2 +\rho^2
d\Omega_3^2}{(1-\frac{(z+a)^2+\rho^2}{R_{ads}^2})^2}
\eeq
The parameter $a$ can be determined from the Israel condition
and equals to
\beq
a=\frac{4T_0}{h_{+}^2 -h_{-}^2}
\eeq
where $T_0$ is the tension of the neutral brane. The brane world
on the neutral brane enjoys $AdS_4$ geometry.

More general bounce in the  $AdS_5$ geometry has been
found in \cite{gs2}. In this case one considers three charged branes
glued together along junction manifold. The branes are considered now
as test objects that it their back reaction on the metric
is neglected.
This solution
is based essentially on the consideration of the minimal charged surfaces in AdS with
the homogeneous field
described in \cite{mms}. They are classified into three
classes: undercharged ones (saturating a sort of BPS inequality
between charge and tension of the brane),
overcharged ones (breaking the BPS inequality) and BPS ones .
Our bounce now is glued out of three pieces of branes
- of BPS one, of undercharged one and of overcharged one.

Consider the metric of AdS as in \cite{mms},
\beq
\label{metric}
ds^2 =  R_{ads}^2 \left(\cosh^2\!\rho \, d\tau^2 + d\rho^2 + \sinh^2\!{\rho} \,
d\Omega_{d-2}^2 \right),
\eeq
where $R_{ads}$ is the anti-de Sitter radius, and assume that
the curvature form $H=dB$ of the $B$-field is proportional to the volume form
with a constant coefficient (flux density). Assuming
the spherical symmetry
of the brane worldsheet, one reduces the effective action
to :
\beq
\label{action2}
S = TR_{ads}^{d-1} \Omega_{d-2}  \int d\tau \left[
\sinh^{d-2}\rho  \sqrt{ \cosh^2\rho + \left({d \rho
\over d \tau}\right)^2 } - q \sinh^{d-1}\rho  \right]
\eeq
where
$\Omega_{d-2}={2 \pi^{{(d-1)\over 2}} \over \Gamma({d-1 \over 2})}$
stands for the  volume of a unit $(d-2)$ sphere
and  $q$ is a constant made out of the flux density, brane charge
and the brane tension. The condition $q=1$ is identified in \cite{mms} as the
BPS one.  The branes with $q<1$ will be referred to as undercharged ones and
those with $q>1$ - as overcharged.

Upon the change of variables,
\beq
\label{variables2}
\tanh \tau = \tan \theta
\eeq
the metric  (\ref{metric}) is put to the form
\beq
\label{metric2}
ds^2 =  R_{ads}^2 \frac{d\tau^2 + d\theta^{2} + \sin^{2}\theta \,
d\Omega_{d-2}^2 }{\cos^{2}\theta}
\eeq

The minimal surfaces for the case  $q<1$
look as follows:
\beq
\label{under}
\cosh\rho = { \sinh \tau_{m} \over \sinh (\tau +\tau_0) }
\eeq
where $\tanh \tau_{m}=q$.

In coordinates (\ref{variables2}) the undercharged surfaces
take the form
\beq
\label{under2}
\cos\theta=\frac{\sinh(\tau + \tau_0)}{\sinh\tau_{m}}.
\eeq
Restriction of the $AdS_{5}$
metric (\ref{metric2}) onto the undercharged surfaces (\ref{over2})
gives the metric of $AdS_{4}$.

In the case  $q>1$
the relevant charged minimal surfaces look as follows:
\beq
\label{over}
\cosh\rho = { \cosh \rho_{m}\over \cosh (\tau +\tau_0) }
\eeq
where  $\tanh \rho_{m} = 1/q$.
In coordinates (\ref{variables2}) the overcharged surfaces
take the form
\beq
\label{over2}
\cos\theta=\frac{\cosh(\tau + \tau_0)}{\cosh\rho_{m}}.
\eeq
Restriction of the $AdS_{5}$
metric (\ref{metric2}) onto the overcharged surfaces (\ref{over2})
gives the metric of the sphere $S_{4}$.
The BPS case (q=1) can be obtained as a limit from either of the cases
above. Corresponding surfaces look as follows :
\beq
\label{bps}
\cos\theta=\frac{1}{z_{0}}e^{\tau}
\eeq
where $z_{0}$ is a constant.
Apparently, restriction of the AdS metric onto these surfaces
gives flat Euclidean metric.

The bounce which describes the tunneling into the
brane world \cite{gs2} is glued out of three pieces - a piece of BPS brane
located along $z=z_{0}$ section,
where the brane world with the flat metric is defined, a piece of undercharged brane, (\ref{under2}),
located above the
BPS brane
and playing a role of one regulator brane, and a piece of overcharged brane,
located below the BPS brane and playing the role of the other regulator brane.
All three pieces are glued along the junction manifold.
The usual junction conditions are the charge conservation and
the tension forces balance.
Apparently, the configuration sketched above has a finite  action
since none of the constituting pieces reaches the AdS boundary ,
hence the tunneling goes with a finite probability which
can be easily computed.

${\bf {4}}$.
To discuss the different possible continuation
surfaces let us comment on their generic properties
discussed in \cite{hartle}. It was shown that to have
so called real geometries one has to impose the
condition of the vanishing external curvature $K=0$ on
the continuation surface and
the vanishing canonical momenta of all fields involved.
In the case of  spontaneous creation of the brane world
the situation is simpler since we can consider
tunneling of the single
brane degree of freedom that is radius of the bubble quasiclassically. On the other hand
gravity
can be treated classically and the emergent geometry
on the brane worldvolume is defined by its embedding. That
is condition on the continuation surface in our example is
vanishing of the canonical momenta $\dot{R}=0$.

Consider the symmetric bounce corresponding to
the spontaneous brane world production in the external field.
The solution involves spherical bubble
formed by the charged brane
expanding in the Minkowski space after the turning
point. The effective Lagrangian describing the tunneling
process depends on the radius of the bubble and reads as
\beq
L=-4\pi TR^2\sqrt{1-\dot{R}^2} + \frac{4\pi  eHR^3}{3}
\eeq
where $T$ is tension of the charged brane, $H$ is the
constant curvature of the external higher-form field
and $e$ is the brane charge.
The canonical Hamiltonian can be easily derived
\beq
H=\sqrt{p^2 +(4\pi TR^2)^2} -  \frac{4\pi eHR^3}{3}
\eeq
where $p$ is the canonical momentum.
The wave function of the universe on the brane  $\Psi(R)$ depends
on the radius of the bubble and obeys
the equation
\beq
\hat{H}\Psi(R)=0
\eeq
The probability of the creation of the bubble can be easily calculated
\beq
w \propto exp(-\frac{9\pi^2 T^3}{g^2H^2})
\eeq
It can be assumed that
the space outside the bubble is flat but is
$AdS_5$ inside the bubble. In this simple
case all directions of time arrow are
equivalent due to the spherical symmetry. The
cut of the bubble which yields the direction
of time arrow can be identified with any section involving the
center of the bubble.

However in  more complicated case of spherically
asymmetric bounces described above   generically there
are  several possibilities to
perform the analytic continuation to the
Lorentzian space. Consider the solution
to the equation of motion found
in  \cite{gs1}. At Fig.1a
the simplest analytic continuation is presented
with the vertical direction of the Lorentzian time.
The process corresponding to this solution looks as follows.
First, two small bubbles are produced in the Euclidean
space then these bubbles "collide"  and form a joint
boundary in the Euclidean region. Later on at
the turning point the configuration is continued
into the Lorentzian space.
The action calculated on the  solution yields the probability
of the process
$$
nothing \rightarrow charged \quad brane + neutral \quad  brane
$$
The configuration
arising in the Lorentzian space looks like
two spherical segments of the charged branes connected
by the flat neutral brane. It is assumed that
brane world is defined  on the neutral brane.
The  neutral brane has $AdS_4$  worldvolume metric and charged brane
provides a kind of regularization.

Another continuation into the
Lorentzian space is possible  with the
vertical direction of time at Fig 1b.
In this case we   fix initial
and final states in the process to be
$$
nothing \rightarrow charged \quad brane.
$$
and asymmetric bounce provides the nonperturbative
correction to the probability defined by
spherically symmetric one.
That is the total probability from two bounces reads as  $w_0 +w_1$ where
$w_0$ is the  contribution prom the spherically
symmetric bounce and  $w_1$ is correction amounted from Fig.1b type
configuration involving S-brane localized at the imaginary time.

Consider more general asymmetric bounce involving multiple
branes. The example
of the corresponding configuration is presented at Fig.2
and the metric between two neighbor branes reads as
\beq
ds^2=\frac{dz^2+d\rho^2 +\rho^2
d\Omega_3^2}{(1-\frac{(z-a_n)^2+\rho^2}{R_{ads}^2})^2}
\eeq
where the position of the  center of n-th neutral brane  equals to
\beq
a_n=(2n-1)a
\eeq
With the continuation surface $A$ we get charged brane with
large number of neutral branes in the final state while the continuation
surface $B$ amounts to the charged brane only.
Evidently the contribution of this solution
to the tunneling probability is suppressed
compared to the single intermediate brane case.
Let us also note that since  S-branes have wrong kinetic
terms corresponding to the fluctuations in the time
direction with the infinite number of S-branes we have a kind of
dynamical mechanism supporting
the "deconstruction of time" scenario \cite{bgk}.

Note the interpretation of  S-branes
in the imaginary time developed in \cite{rastelli,llm,msy}.
According to this interpretation the infinite chain of
S-branes can be related to the closed string mode of the
decaying brane located at the fixed time. It has been proven in the
stringy framework that the  background
of the S-branes in the imaginary time is equivalent to the
additional boundary state  on the worldsheet.
Equivalently one could introduce
the tachyon potential
\beq
\delta L= \lambda coshX_0
\eeq
where $\lambda$ fixes the distance between the S-branes
in the imaginary time direction
\beq
\lambda\propto \delta t_{E}
\eeq
and the lifetime of the decaying brane.
The  analysis of \cite{rastelli} implies that the relation
with the closed string modes survives for the finite
number of S-branes distributed irregularly in the imaginary
time. The value of $\lambda=1/2$ corresponds to the special
case when S-branes provide the Dirichlet boundary conditions
for the open strings.

Comparing this picture with our model case we conclude that array of
the neutral S-branes in the imaginary time can be identified
with the decaying closed string.
Note that to justify the quasiclassical calculation we have
to assume that the radii of the corresponding bubbles are
large enough. Hence it is naturally to expect that boundary effects
on the neutral S-branes don't change this interpretation.
Let us remark that  idea to use  decaying brane for the generation of
matter in the cosmological framework has been discussed
in \cite{leigh}.

The wave function
of the universe depends on the size of the
continuation surface which simultaneously provides
the quasiclassical metrics on the worldvolume of the brane.
If we consider continuation surface of the type B at Fig 2. the
total  wave function of the universe of this  type looks as
\beq
\Psi_{\infty} =\sum_{k=0}^{\infty} a_k \psi_k
\eeq
where $\psi_k$ corresponds to the wave function with
$k$ S-branes in the imaginary time involved.
The total probability of the quantum creation of the
brane world in this channel is related to $\Psi_{\infty}$ as
follows
\beq
e^{-S}\propto <\Psi_{\infty}|\Psi_{\infty}>
\eeq
and different terms in the scalar product correspond to the
different continuation surfaces into the Lorentzian space-time.
For instance, two different horizontal cuts
at Fig.1b correspond to the contributions
$<\Psi_{0}|\Psi_{1}>$ and $<\Psi_{1}|\Psi_{0}>$
to the probability of the process. Just as in two-dimensional toy
example if we make two horizontal cuts simultaneously
the emerging object can be identified with the density
matrix of the brane universe.

For the continuation surfaces of the type A at Fig.2 the total wave function
of the universe can be defined as the infinite sum as well
\beq
\Phi_{\infty}=\sum_{k=0}^{\infty} b_k \phi_k
\eeq
however with the different interpretation. If the
brane world is defined on the charged brane then $\phi_{k}$
corresponds to the contribution to the wave function involving
k additional neutral branes. Contrary if the brane world
is defined on stack of the neutral branes then
charged branes play the role of additional matter. Note
that we have the simple relation
\beq
 <\Psi_{\infty}|\Psi_{\infty}>= <\Phi_{\infty}|\Phi_{\infty}>.
\eeq
whose counterpart exists for each particular asymmetric bounce.

\begin{figure}
\epsfxsize=10cm
\centerline{\epsfbox{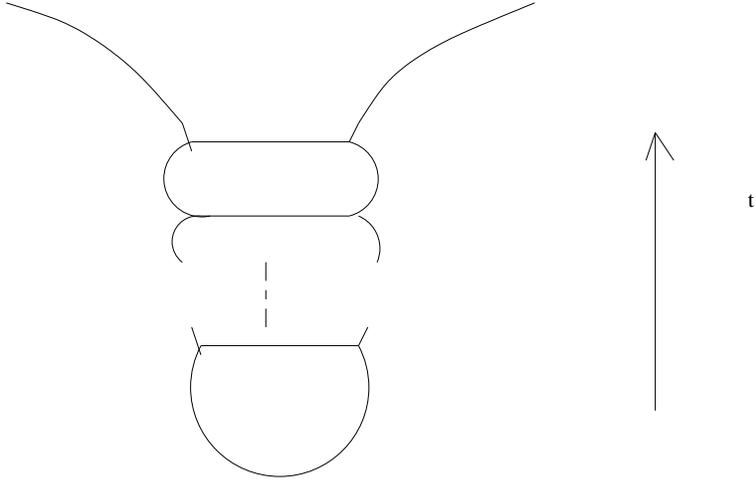}}
\caption{\small The schematic representation
of the contribution to the total  Hartle-Hawking wave function
$\Psi_{\infty}$ involving
the array of S-branes in the imaginary time.
}
\label{fig:124}
\end{figure}

The similar analysis can be done for the spherically
asymmetric gravitational bounce discussed in \cite{tye, brushtein, bk}.
In this case there are several inequivalent continuation surfaces
as well which differ by the  different values of the scale
factors in the emerging universes.

${\bf {5}}$. In this note we have argued that generically
spherically asymmetric bounce
describing nucleation of the brane world admits several
inequivalent continuations to the Lorentzian space-time with the
different directions of the time arrow and Hartle-Hawking
wave functions corespondingly. Contrary to the asymmetric
gravitational bounce in \cite{bk} we treat "matter"
quasiclassically implying the classical gravity.
We analyzed the simplest example of
nucleation of the brane world in the nondynamical four-form
field with the constant curvature. The existence
of the neutral and charged branes with respect to the external field
was assumed in our model example.
In this case there are two essentially different continuations
into the Lorentzian signature space. One continuation yields brane
world on the expanding charged brane with the positive cosmological
constant. The second continuation provides the brane world
on the neutral brane or array of neutral branes at rest with
negative cosmological constant. To some extent the time variable
in one continuation can be identified with the RG scale
in  another one.

The natural question concerns the most preferable final
state of the nucleation process.  It seems that it is
natural to consider all possible continuation
surfaces on the equal footing
hence one has to sum over brane worlds emerged. Probably
some entropy factors could be essential due to the different
number of branes in the final state .

Another point of concern deals with the multiple
asymmetric bounces. In the case of spherically
symmetric Euclidean bounces summation over the number
of bounces does not amount to the selection problem
for the direction of the time arrow. However
in asymmetric case the situation is more subtle.
Consider for example the Euclidean solution
involving two asymmetric bounces with different orientations
of the neutral branes. It is impossible
to consider the same continuation surface
for them since they have different turning points.
Hence one could consider more complicated continuation
surface in the Euclidean space or take into
account the interaction between the bounces. The interaction
of the differently oriented bounces seems to be minimized
when neutral branes are parallel however this point
needs for the separate analysis.

In this note we considered the bounce
built from the set of branes with the
same dimension, say 3-branes providing four-dimensional
brane world. However the bounces
with the different brane content could be considered
as well. For instance, we can consider the four-rank field
creating the 3-brane spherically asymmetric  bubble
with the set of strings inside. In this case in one
"continuation channel" we get four dimensional brane world
on 3-brane worldvolume. In the different "channel" we get
single 3-brane and  set of strings. Oppositely one can consider
higher rank form fields creating 5-branes. The asymmetric
bounce could be obtained from the bubble of 5-brane with the
number of 3-branes inside. Then in one channel we can
consider the creation of five-dimensional brane world while
in the second channel the creation of the brane world on 5-brane
or 3-branes.

Finally let us note that the opposite process of the
brane world decay in the external field is possible.
To this aim consider the neutral 3-brane in the external
four-four field with the constant curvature. The
decay can occur due to the Euclidean bounce configuration
corresponding to the induced Schwinger process in higher
dimensions similar to the processes considered in \cite{gs3}.
It can be considered as an example of the generic
Myers phenomena in the external fields \cite{myers}.
From the worldvolume point of view the process of decay
looks as the blow up of the four dimensional brane world
into the fifth dimension.

I am grateful to A. Barvisnky and A.Rosly for the useful comments.
The work was supported in part by grants CRDF RUP2-261-MO-04
and RFBR-04-011-00646.

\end{document}